\documentclass{research4cacm}
\usepackage{multirow}
\usepackage[hyphens]{url}
\usepackage{hyperref}
\hypersetup{breaklinks=true}

\usepackage{cite}

\usepackage{enumitem}
\usepackage[utf8]{inputenc}
\usepackage{epsfig,endnotes}
\usepackage{varwidth}
\usepackage{algorithm}
\usepackage{algpseudocode}
\usepackage{balance}
\usepackage{nicefrac}
\usepackage{amsmath}
\usepackage{mathtools}
\usepackage{multirow}
\usepackage{bigdelim}
\usepackage{mathtools}
\usepackage{amssymb}
\usepackage{indentfirst}
\usepackage{booktabs}
\usepackage{enumitem}
\usepackage{float}
\usepackage{graphicx}
\usepackage{caption, subcaption}
\usepackage[normalem]{ulem}
\usepackage[titletoc]{appendix}
\usepackage{flushend}
\usepackage{subcaption}
\usepackage[normalem]{ulem}
\usepackage{xpatch}
\usepackage{stmaryrd}
\usepackage{MnSymbol}
\usepackage{xspace}
\usepackage{amssymb}
\usepackage{framed,mdframed}
\usepackage[dvipsnames]{xcolor}
\usepackage{hyperref}
\usepackage{csquotes}
\usepackage{footnote}
\makesavenoteenv{tabular}
\makesavenoteenv{table}

\newcommand{\squishlist}{
	\begin{list}{$\bullet$}
		{
			\setlength{\itemsep}{0pt}
			\setlength{\parsep}{3pt}
			\setlength{\topsep}{3pt}
			\setlength{\partopsep}{0pt}
			\setlength{\leftmargin}{1.5em}
			\setlength{\labelwidth}{1em}
			\setlength{\labelsep}{0.5em} } }
	
	\newcommand{\squishend}{
\end{list}  }

\newcommand{\eat}[1]{}

\graphicspath{{./figs/}}
\DeclareGraphicsExtensions{.pdf,.jpg,.png}

\makeatletter
\def\@copyrightspace{\relax}
\makeatother

\begin{document}
\newcommand{\myitem}[1]{\vspace*{0.02in}\noindent\textbf{#1}}

\title{Securing Internet Applications from Routing Attacks}
\author{
{Yixin Sun}\\
University of Virginia
\and
{Maria Apostolaki}\\
ETH Zurich
\and
{Henry Birge-Lee}\\
Princeton University
\and
{Laurent Vanbever}\\
ETH Zurich
\and
{Jennifer Rexford}\\
Princeton University
\and
{Mung Chiang}\\
Purdue University
\and
{Prateek Mittal}\\
Princeton University
}

\maketitle
\begin{abstract}
Attacks on Internet routing are typically viewed through the lens of
availability and confidentiality, assuming an adversary that either
discards traffic or performs eavesdropping. Yet, a strategic
adversary can use routing attacks to compromise the security of
critical Internet applications like Tor, certificate authorities,
and the bitcoin network. 

In this paper, we survey such application-specific routing attacks and argue that both application-layer and
network-layer defenses are essential and urgently needed. 
While application-layer defenses are easier to deploy in the short term, we hope that our work serves to provide much needed momentum for the deployment of network-layer defenses.
\end{abstract}

\section{Introduction}

The Internet is a ``network of networks" that interconnects tens of thousands of separately administered networks.  The Border Gateway Protocol (BGP) is the glue that holds the Internet together by propagating information about how to reach destinations in remote networks.  However, BGP is notoriously vulnerable to misconfiguration and attack.  The consequences range from making destinations unreachable (e.g., Google's routing incident caused widespread Internet outage in Japan~\cite{google2017}), to misdirecting traffic through unexpected intermediaries (e.g., European mobile traffic routed through China Telecom due to improper routing announcements from a Swiss datacenter~\cite{chinatelecom2019}), to impersonating legitimate services (e.g., traffic to an Amazon DNS server rerouted to attackers who answered DNS queries with fraudulent IP addresses~\cite{etherwaller}).
Efforts to secure the Internet routing system have been underway for many years~\cite{rpki,bgpsec,boldyreva2012provable,gill2011let,lychev2013bgp,kent2000secure}, but the pace of progress is slow since many parties must agree on solutions and cooperate in their deployment.

\begin{figure}
\centering
\includegraphics[width=0.9\linewidth]{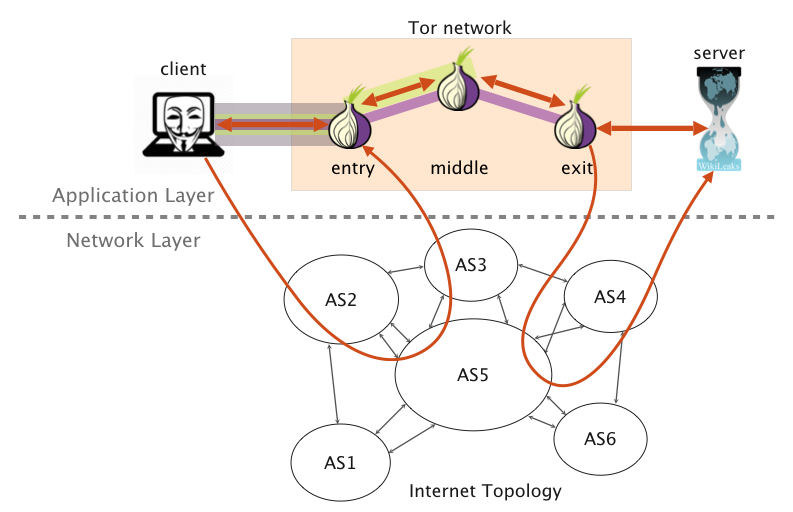}
\caption{BGP routing affects who can observe Tor traffic.}
\label{fig:cross_layer}
\end{figure}

In the meantime, more and more users rely on the Internet to access a wide range of services, including applications with security and privacy concerns of their own.  Applications such as Tor (The Onion Routing) allow users to browse anonymously, certificate authorities provide certificates for secure access to web services, and blockchain supports secure cryptocurrencies.  However, the privacy and security properties of these applications depend on the network to deliver traffic; Figure~\ref{fig:cross_layer} illustrates the \emph{cross-layer} interaction between Tor and the underlying network.  Application developers abstract away the details of Internet routing, but BGP does not provide a sufficiently secure scaffolding for these applications. This gap leaves the vulnerabilities due to routing insecurity significantly underestimated. 
While routing attacks are well known, they have been viewed primarily as affecting availability (when misdirected traffic is dropped) and confidentiality (when data is not encrypted). This paper provides a new perspective by showing that routing attacks on Internet applications can have even more devastating consequences for users---including uncovering users (such as political dissidents) trying to communicate anonymously, impersonating websites even if the traffic uses HTTPS, and stealing cryptocurrency.

This paper argues that the security of Internet applications and the network infrastructure should be considered together, as vulnerabilities in one layer lead to broken assumptions (and new vectors for attacks) in the other.  We first give an overview of routing security. %
Then, we discuss how cross-layer interactions enable routing attacks to compromise popular applications like Tor, certificate authorities, and the bitcoin network. 
Given the slow adoption of secure routing solutions, we discuss how applications can take into account the underlying routing properties and employ application-layer defenses to mitigate routing attacks. 
We believe that application-layer and network-layer solutions are interconnected and both are essential to secure Internet applications. 
While application-layer defenses are more easily deployable, we hope to motivate the community to redouble efforts on secure routing solutions and tackle BGP's many security problems once and for all.

\section{Routing Attacks}
\label{sec:background}

Routing attacks occur in the wild, and are getting increasingly prevalent and more sophisticated. 
We dissect routing attacks from the perspective of an attacker, and review existing defenses. %
In particular, the ability to divert targeted traffic via routing attacks is an emerging threat to Internet applications. We further demonstrate how routing attacks compromise three applications in Sections~\ref{sec:tor},~\ref{sec:pki}, and~\ref{sec:bitcoin}.

\subsection{How BGP Works}
\label{ssec:preference}
The Internet consists of around 67,000 Autonomous Systems (ASes)~\cite{cidr-report}, each with an AS number (ASN) and a set of IP prefixes.%
Neighboring ASes exchange traffic in a variety of bilateral relationships that specify which traffic should be sent and how it is paid for.  Such agreements can generally be classified into two types: a \emph{customer-provider} relationship, where the customer pays the provider to send and receive traffic to and from the rest of the Internet, and a \emph{peer-to-peer} relationship, where no money is exchanged but traffic must be destined for the peer or its customers.

\begin{figure}
\centering
\includegraphics[width=.9\linewidth]{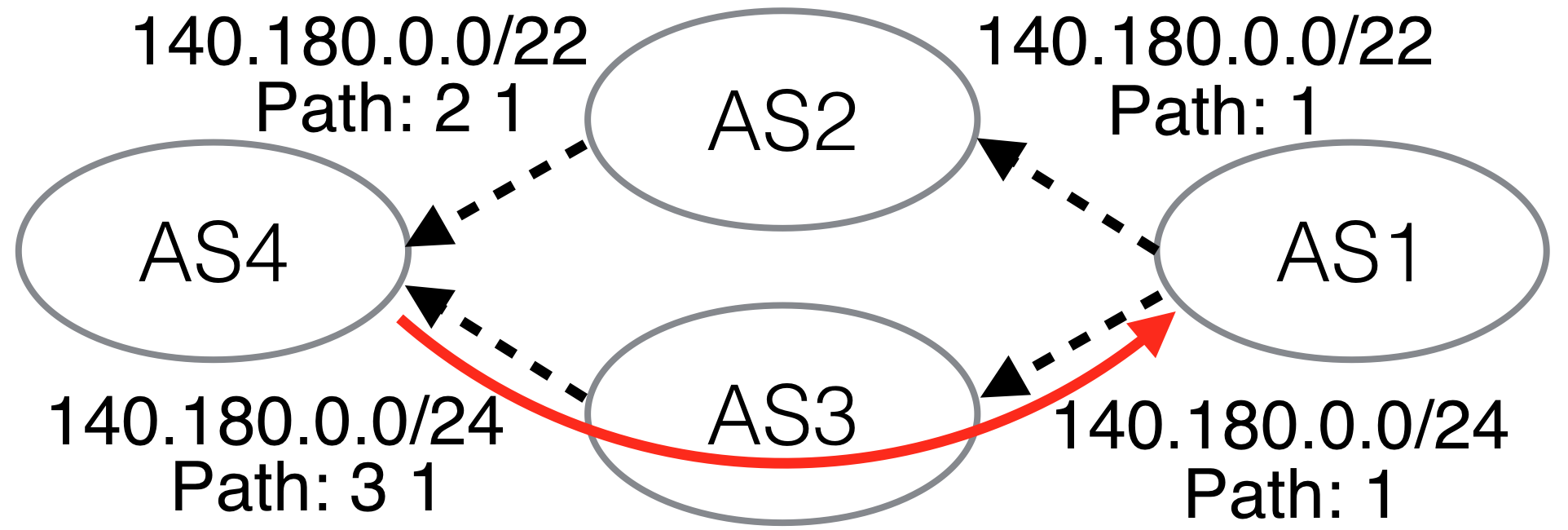}
\caption{AS4 routes traffic to AS1 via AS3 for destination IPs within $140.180.0.0/24$ based on longest matching prefix.}
\label{fig:as_announce}
\end{figure}

Routing among the ASes is governed by the Border Gateway Protocol (BGP), which computes paths to 
destination prefixes. 
ASes choose one ``best'' route to a prefix based on a list of factors, with the top two generally being: 
(1) Local Preference: a path via a customer is preferred over path via a peer,
which is preferred over a provider; (2) Shortest Path:
a path with the fewest AS hops is preferred.
The AS will then add the route into its local \emph{Routing Information Base}, 
and further propagate the route to its neighbors based on routing policies after prepending itself in the path.

ASes forward packets using the path to the longest matching prefix of the destination IP.
In Figure~\ref{fig:as_announce}, AS1 announces $140.180.0.0/22$ via neighbor AS2, and $140.180.0.0/24$ via neighbor AS3. %
AS4 forwards packets to $140.180.0.0/24$ via AS3 based on the longest prefix match. 
Note that, in general, the longest prefix that can be successfully propagated is $/24$; many
ASes filter prefixes that are longer than $/24$ by default. 

\subsection{Goals of Routing Attacks}

By default, ASes trust routing announcements from other ASes. 
Routing attacks happen when an AS announces an incorrect path to a prefix, 
causing packets to traverse through and/or arrive at the attacker AS. 
We discuss the goals of the attacker from two perspectives: \emph{whom to affect} and \emph{what to achieve}. 

\subsubsection{Whom to Affect}

Routing attacks affect two groups of victims: (1) destinations, whose prefixes are announced by the attacker, and (2) senders, who send packets to the attacked prefixes. 

{\bf Destinations.} %
YouTube was the targeted destination of a hijacking incident in 2008, where Pakistan authorities tried to block access to YouTube~\cite{youtube2008}. Pakistan Telecom (AS17557) announced the prefix $208.65.153.0/24$, which was a subnet of $208.65.152.0/22$ announced by YouTube (AS36561).

{\bf Senders.} The attacker can divert global traffic from all senders on the Internet, or selectively target only traffic from certain senders. %
In the YouTube incident above, the goal was to target only senders within Pakistan; however, the attack unintentionally affected all senders around the globe. 

\subsubsection{What to Achieve}

Historically, the most visible effect of routing attacks is \emph{outage}, 
where attackers drop packets and make the destinations unreachable. 
This type of attack that ``blackholes" the traffic is also characterized as a \emph{hijack attack}.
However, the attacker's goals can be more sophisticated.

{\bf Surveillance.} Authorities may use routing attacks to perform surveillance and target traffic from senders in certain regions. Intelligence agencies such as NSA could launch routing attacks to make certain traffic easier to intercept for surveillance~\cite{nsashaping}. Traffic from the targeted region would be rerouted to the authorities, who forward the traffic to the destinations while monitoring the activities. This type of attack is usually characterized as an \emph{interception attack}, where the legitimate destinations still receive the traffic. 
Interception attacks are much harder to notice than hijack attacks since they do not interrupt the communication, though performance may degrade due to more circuitous paths. 
Furthermore, authorities could exploit routing attacks to surpass legal restrictions
by diverting domestic traffic (e.g., emails between Americans) to foreign jurisdictions to conduct surveillance~\cite{goldberg2017nsa}.

{\bf Impersonation.} Attackers can impersonate destinations to deceive the senders
by intercepting packets via either hijack or interception attacks and replying with forged responses. 
These attacks can have damaging consequences. In 2018, attackers used routing attacks to impersonate Amazon's authoritative DNS service~\cite{etherwaller} and answered DNS queries for a cryptocurrency website with Russian IP addresses. The users were then directed to a fraudulent site which they believed was their real cryptocurrency service. Consequently, cryptocurrency was stolen. 
Attackers may also impersonate large number of IP addresses to originate spam or other malicious traffic~\cite{bitcanal2018}.

{\bf Cross-layer attacks on applications}. Attackers may further exploit the diverted traffic to perform more sophisticated attacks on networked systems and applications. The specific goals vary depending on the functionalities of the applications. In this paper, we demsontrate routing attacks on three applications: deanonymizing Tor users via traffic analysis on the Tor network (Section~\ref{sec:tor}), obtaining bogus digital certificates for websites from certificate authorities (Section~\ref{sec:pki}), and preventing blockchain systems from reaching consensus (Section~\ref{sec:bitcoin}).

\subsection{Attack Methodology}

Attackers need to decide (1) which prefix to announce, (2) which path to announce, and (3) which ASes should receive the announcement.

\subsubsection{Which Prefix to Announce}
\label{subsec:prefix_announce}

Attackers can announce either (1) a sub-prefix (i.e., more-specific prefix) of the target prefix, or (2) an equally-specific prefix same as the target prefix. Note that a less-specific prefix would not be used in packet forwarding and hence would not constitute a successful attack. 

{\bf Affecting global traffic by announcing sub-prefixes.} 
Since forwarding is based on longest prefix match, sub-prefix attacks are highly effective at hijacking traffic from all senders.  However, since most ASes filter announcements for prefixes longer than $/24$, sub-prefix attacks on $/24$ prefixes would not be effective.

\begin{figure}
\centering
\includegraphics[width=.9\linewidth]{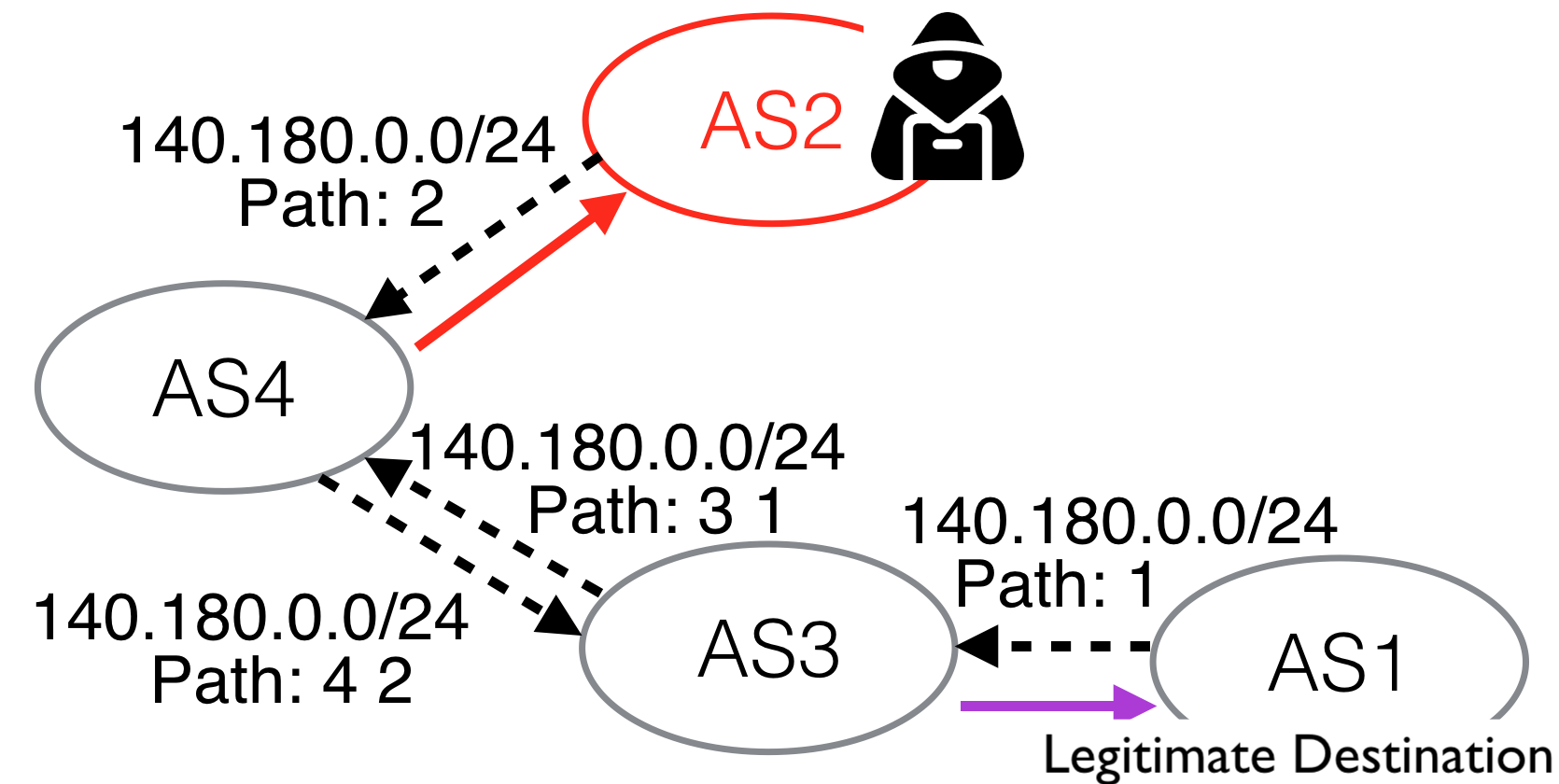}
\caption{AS2 (attacker) announces an equally-specific prefix as AS1 (legitimate destination). AS4 prefers the path to AS2, while AS3 prefers the path to AS1. Only AS4 is affected by the attack.}
\label{fig:equal_specific}
\end{figure}

{\bf Targeting selective traffic by announcing equally-specific prefixes.} %
An AS that receives both the legitimate announcement and the attacker's announcement would pick one based on routing preferences.
Note that some ASes may only receive one announcement.
In Figure~\ref{fig:equal_specific}, AS2 (attacker) announces the same $/24$ prefix as the destination AS1, and AS4 prefers the path to AS2 while AS3 still prefers the path to AS1. 
This attack generally affects only parts of the Internet and does not have global impact. 
However, it is stealthier due to its local impact and enables targeted attacks on certain senders.

\subsubsection{Which Path to Announce}
\label{subsec:path_announce}

The attacker may put itself as the origin of the prefix, which naturally constitutes a hijack attack. %
Yet, a more sophisticated attacker has a range of other options.

{\bf Evading detection by forging the victim AS.}  The attacker can add the legitimate destination AS to the end of its path, so the announcement has the same ``last hop'' (i.e., ``origin'') AS as a legitimate announcement.  This makes the attack stealthier since some defenses (e.g., monitoring systems and RPKI) only check the origin AS of the announcement instead of the full path. Note that the path now appears one hop longer, which may reduce the number of ASes that pick the attacker's route over the legitimate route.

\begin{figure}
\centering
\includegraphics[width=.9\linewidth]{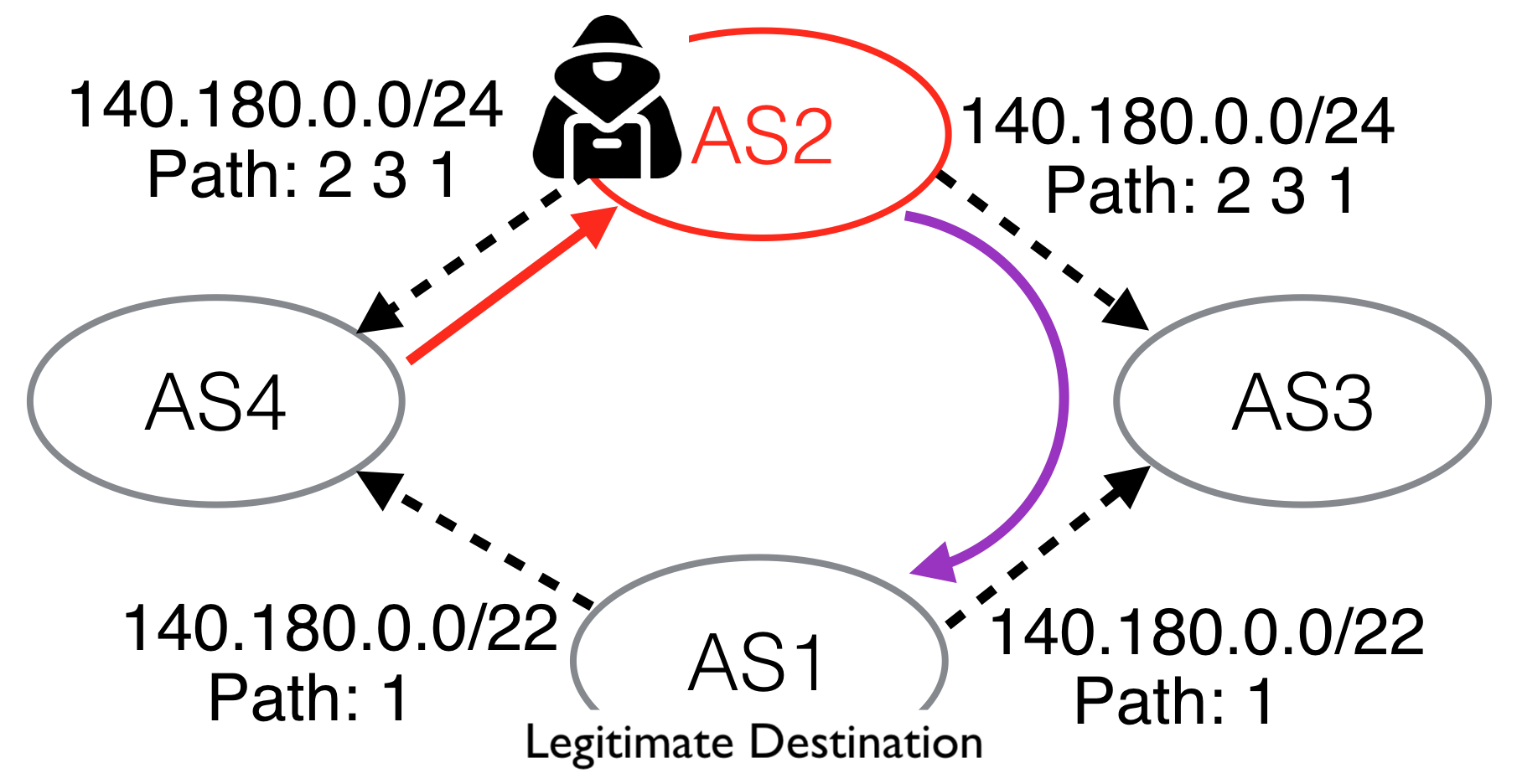}
\caption{AS2 (attacker) ``poisons" the path by appending AS3 and AS1 (legitimate destination) in the path, which preserves a legitimate route from AS2 to AS1.}
\label{fig:path_poison}
\end{figure}

{\bf Interception attack via AS path poisoning.} A sophisticated attacker can append a set of carefully-selected ASes at the end of the path. 
These ASes should constitute a legitimate path from the attacker to the destination AS. 
The appended ASes will ignore the attacker's announcement because of BGP loop prevention, 
which consequently helps preserve legitimate routes from the attacker AS to the destination AS. 
This attack is known as the ``AS path poisoning attack" (Figure~\ref{fig:path_poison}). 
This attack is very stealthy and effective at performing interception attack while announcing a sub-prefix. 

\subsubsection{Which ASes Should Receive the Announcement}
\label{subsec:as_announce}

Instead of sending the announcement to all neighbors, 
a strategic attacker may attempt to control who can receive the announcement 
to increase attack stealthiness, perform an interception attack, or target certain senders. 
We discuss two techniques to limit announcement propagation. %

{\bf Announcing to certain neighbors.} 
Attackers may exploit routing policies to control attack propagation by only announcing to certain peers and customers. 
These announcements will only be propagated ``down" to the peer's customers, but not to its providers. Consequently, only selected ASes will hear the announcements. 

{\bf BGP communities.} 
BGP communities are optional attributes that can be added to an announcement to control routing policies in upstream ASes, for purposes such as traffic engineering.
Attackers may exploit BGP communities to strategically control attack propagation such that selected ASes will never hear or will not prefer the bogus announcements, 
and thus increase the effectiveness and viability of interception attacks~\cite{birge2019sico}. \\

\subsection{Routing Defenses}
\label{subsec:defenses}

Defending against routing attacks is challenging due to the lack of ``ground truth'' to inform  whether a path is ``correct". Seemingly suspicious announcements could be legitimate paths used by ASes to optimize network performance. 
Many solutions have been proposed that rely on different sources of information as ``ground truth".

{\bf Anomaly detection via BGP monitoring.} 
BGP monitoring systems detect anomalous routing announcements by using historical routing data to infer
the ``expected'' origin ASes or paths for prefixes~\cite{lad2006phas, zhang2008ispy, zheng2007light, hu2007accurate, shi2012detecting, qiu2007detecting}.
They typically do not require changes to the routing protocol and hence are highly deployable. 
However, many early efforts on monitoring systems focused on catching ``easy'' attacks (e.g., mismatched origin ASes), but failed to detect more sophisticated attacks such as interception attacks. 
Furthermore, relying on historical data to infer ground truth is prone to false positives (flagging legitimate routes) and false negatives (missing real attacks). 

{\bf Defensive filtering via preset knowledge.} 
ASes often perform prefix filtering on announcements received from direct customers. 
It is effective against attacks launched by customer ASes, 
but does not prevent ASes from attacking their direct or indirect customers. 
A more advanced filtering technique is AS path filtering, 
which uses a whitelist of paths for announcements received from peering ASes 
based on prior information exchange~\cite{peerlocking}. 
It extends the knowledge base further from the sole knowledge of an individual provider on its customers (as in prefix filtering), to a collective knowledge base exchanged and built among a network of trusted peers.
The MANRS project~\cite{manrs} has outlined best practices for using filtering techniques to protect the routing infrastructure. 

{\bf Origin validation.} 
The Resource Public Key Infrastructure (RPKI) is a public key infrastructure that stores cryptographic attestations, known as Route Origin Authorizations (ROAs), indicating which ASes are authorized to originate which prefixes~\cite{rpki}. 
Upon receiving an announcement, ASes perform Route Origin Validations (ROV) to filter routes originated from invalid ASes.
RPKI utilizes cryptographic primitives to make the knowledge base available to \emph{all ASes} as opposed to only direct neighbors in defensive filtering. 
Even though ROV only validates the origin AS instead of the full path, it can already be effective at preventing many attacks. 
However, currently less than 20\% of the prefixes have valid ROAs~\cite{nist-rpki} and even fewer ASes are correctly performing ROV~\cite{reuter2018towards}. 

{\bf Path validation.}
BGPsec uses cryptographic primitives to validate the \emph{whole AS path}~\cite{bgpsec}. 
It is an \emph{online} protocol, %
as opposed to a separate offline lookup (like ROV). 
Each AS in the path generates a cryptographic signature which is added to the path as the announcement propagates through the network. 
While BGPsec provides validation of the full path, it places a heavy burden on BGP routers. %
It also requires all ASes along a path to participate, making incremental deployment challenging.
We have yet to see real-world deployment of BGPsec.

In this paper, we provide a new angle into building defenses --- in addition to network-layer defenses, applications can build their own application-layer defenses by taking into account the underlying routing properties. 
We also highlight the importance of deploying defenses against sophisticated attacks, which are stealthier and effective at compromising Internet applications.

\section{The Tor Network}
\label{sec:tor}

Tor is the most widely used anonymity system~\cite{dingledine2004tor}. It carries terabytes of traffic every day and serves millions of users~\cite{tormetrics}. 
However, network-level adversaries can deanonymize Tor users 
by launching routing attacks to observe user traffic and subsequently performing correlation analysis. 
Furthermore, the attacks have broad applicability to low-latency anonymous communication systems beyond Tor (e.g., I2P anonymous network or even VPNs).

\subsection{How Tor Works}

\begin{figure}
\centering
\includegraphics[width=.9\linewidth]{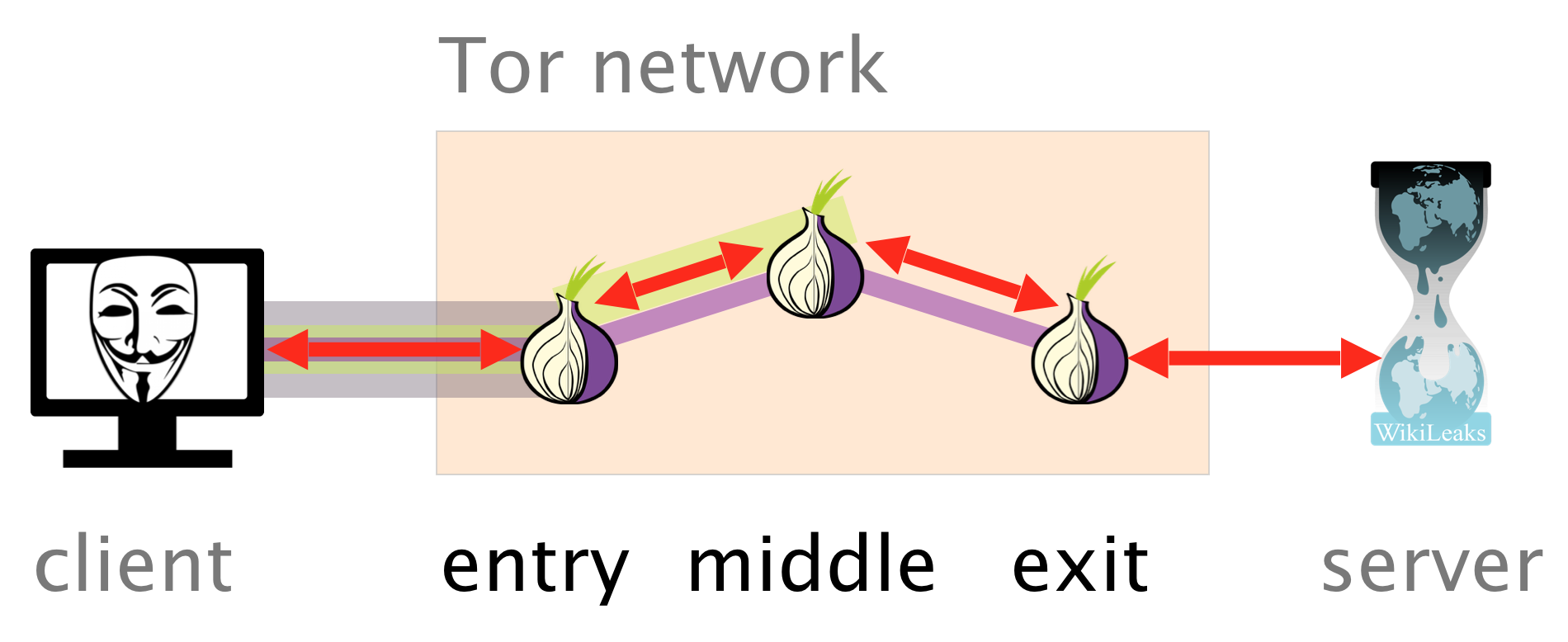}
\caption{The Tor network.}
\label{fig:tor_background}
\end{figure}

To prevent an adversary 
from associating a client with a destination server, Tor encrypts the
network traffic and sends it through a sequence of \emph{relays} (proxies) before going to the destination. 
The client selects three relays (entry, middle, exit), 
and constructs a \emph{circuit} through them with \emph{layered encryption}
by repeatedly encrypting the next hop with the keys of the current hops (Figure~\ref{fig:tor_background}).
Each relay only learns the previous and next hops, and no relay or local network observer can identify both the source and destination. 

However, Tor is known to be vulnerable to network-level adversaries who can observe traffic at both ends of the communication, i.e., between client and entry, and between exit and server. 
By default, Tor does not obfuscate packet timings, so the traffic entering and leaving Tor are highly correlated. 
An adversary on the path at both ends can then perform traffic correlation analysis on the packet traces to deanonymize the clients. 

\begin{figure*}
 \centering
 \begin{subfigure}[t]{0.9\columnwidth}
 \includegraphics[width=\columnwidth]{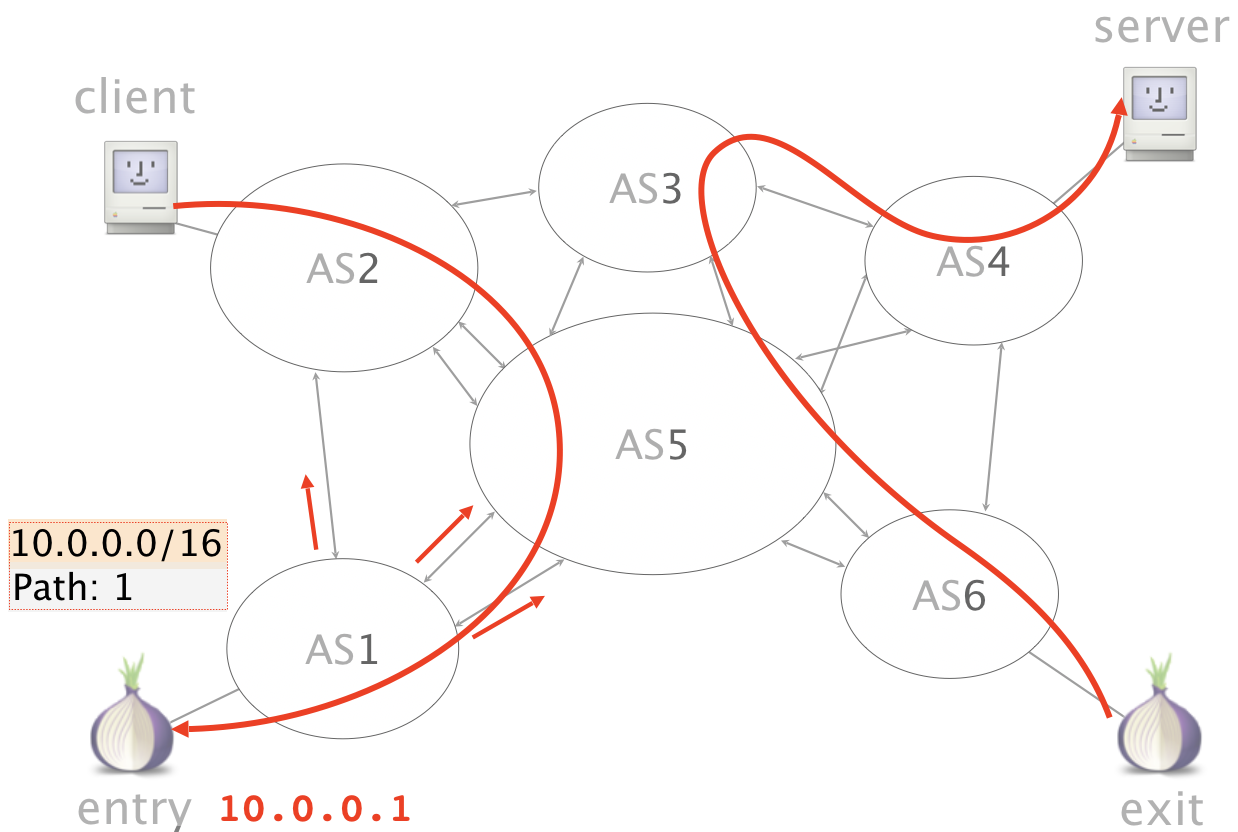}
 \label{fig:hijack_left}
 \end{subfigure}
 \begin{subfigure}[t]{0.9\columnwidth}
 \includegraphics[width=\columnwidth]{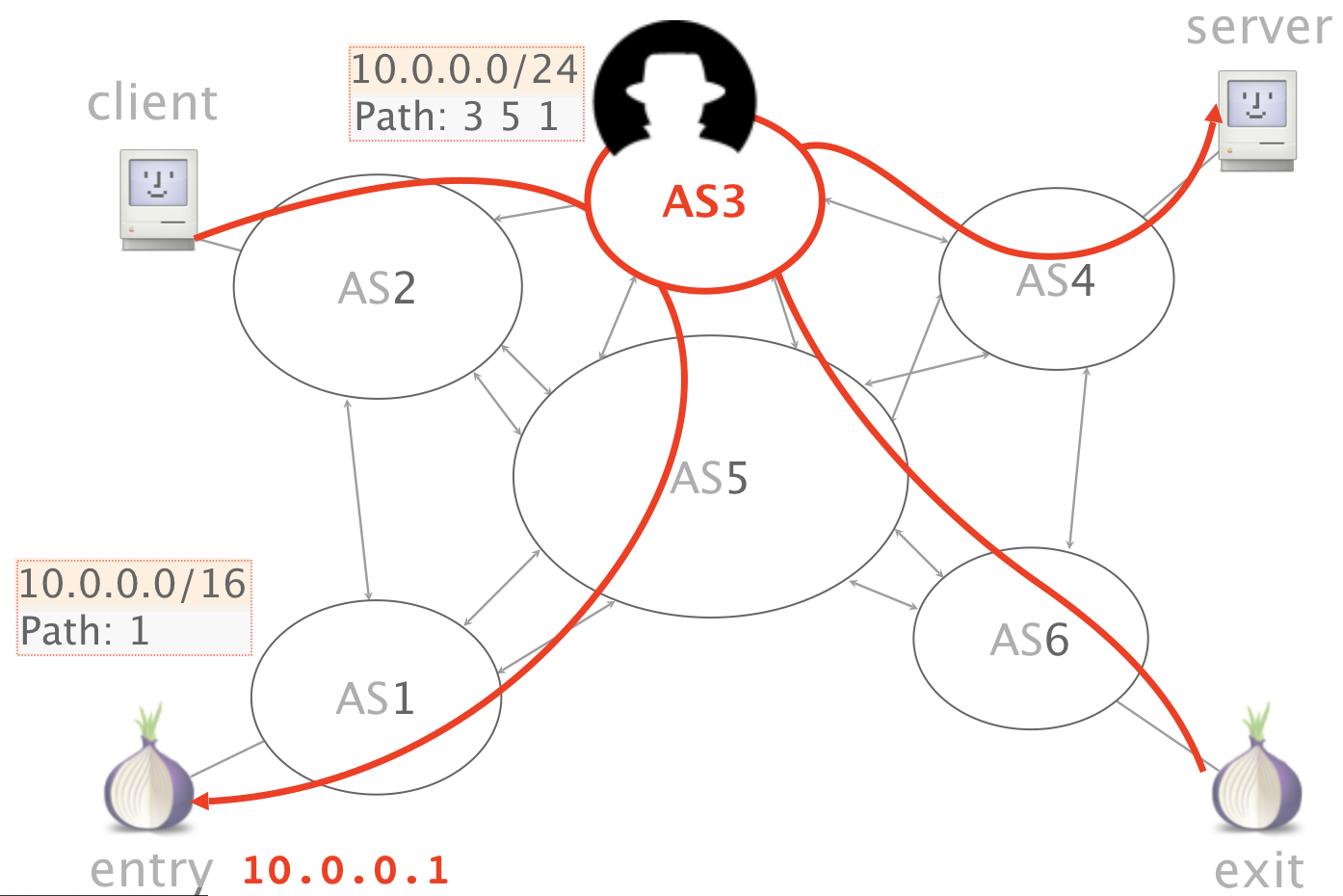}
 \label{fig:hijack_right}
 \end{subfigure}
 \caption{An adversary (AS3) launches an interception attack on the entry relay in AS1 and consequently observes the client-entry traffic in addition to exit-server traffic.}
 \label{fig:interception}
\end{figure*}

\subsection{Routing Attacks on Anonymity Systems}

Traditional attacks from network-level adversaries focus on passive adversaries who are already on the  paths to observe Tor traffic. However, adversaries can exploit active routing attacks to strategically intercept Tor traffic, enabling on-demand and targeted attacks~\cite{sun2015raptor}.

Figure~\ref{fig:interception} illustrates the attack. AS3 (adversary) only sees 
traffic between the exit and the web server, and needs to intercept the traffic between the client and the entry relay. 
It also needs to keep the connection alive in order to capture sufficient traffic for the correlation analysis, i.e., perform an \emph{interception attack}.
AS3 announces an equally-specific prefix of the target prefix which covers the entry relay, while maintaining a valid path (via AS5) to the victim AS1. Consequently, traffic from the client gets routed to the adversary AS3, which forwards the traffic to AS1 to keep the connection alive. 
Similar attacks can be performed to intercept the exit-server connection as well, if the adversary is not already on the path. 

\begin{figure}
\centering
\includegraphics[width=.9\linewidth]{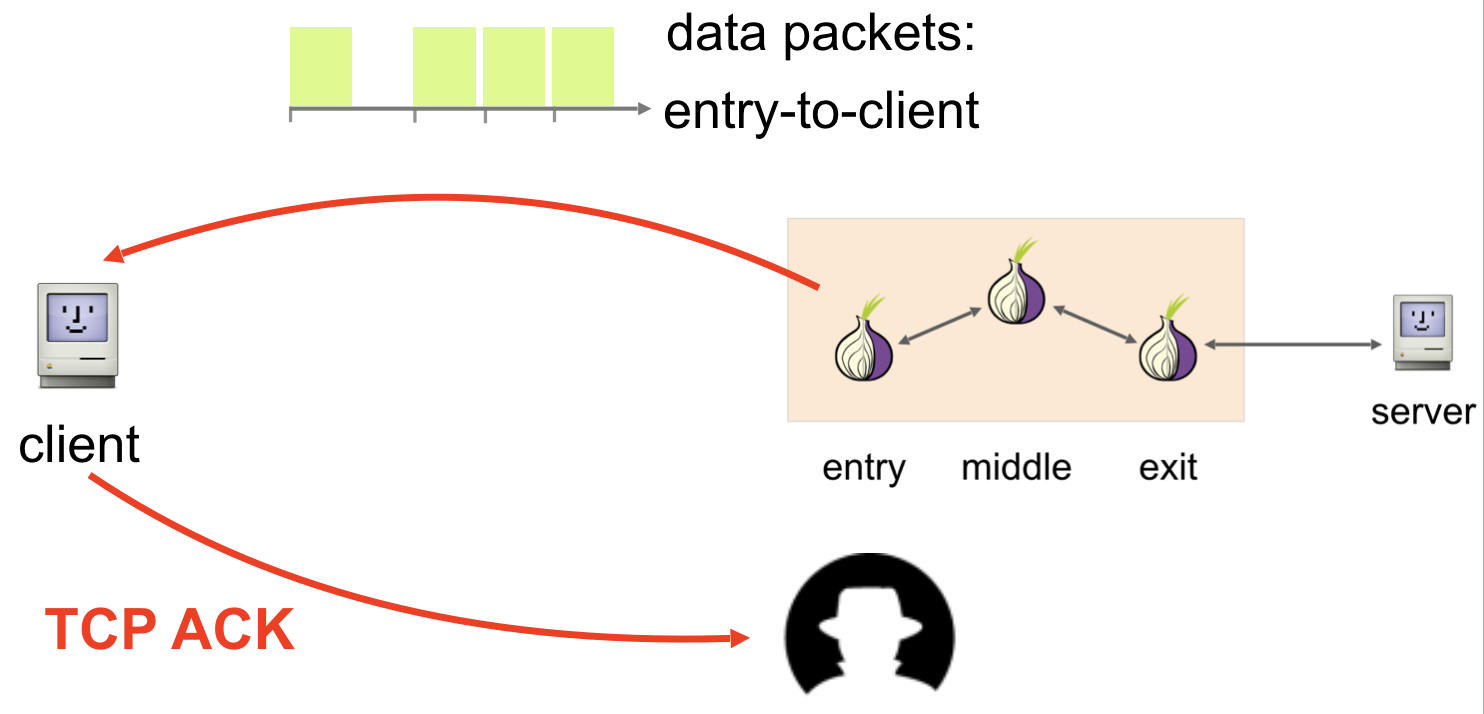}
\caption{The adversary may only see one direction of the traffic but can still perform asymmetric traffic analysis to deanonymize users.}
\label{fig:tor_asymmetric}
\end{figure}

The attacks become more threatening given that seeing \emph{either direction of the traffic} is sufficient, which opens the door to more adversaries. 
Figure~\ref{fig:tor_asymmetric} illustrates the scenario where the user downloads a file from the web server. The adversary performs an interception attack on the entry relay and only sees one direction of the traffic (client to entry relay), which are mostly TCP ACK packets. 
The adversaries then use the sequence and acknowledgment numbers from the TCP header (unencrypted) to determine the sizes of the data packets traveling in the other direction.

The attack was successfully demonstrated on the live Tor network (ethically), by having 50 Tor clients download files from 50 web servers via an entry relay under a prefix controlled by the researchers~\cite{sun2015raptor}. Routing announcements were propagated through the PEERING testbed~\cite{peering19}, and an interception attack was launched on the prefix covering the entry relay. No real user was affected during the attack. 
The attack deanonymized 90\% of the clients in less than five minutes.

\subsection{Defenses to Protect Anonymity}

Many existing defenses cannot sufficiently detect or prevent such interception attacks. Recent works have proposed application-layer defenses for Tor~\cite{sun2017counter,tan2016data}. 

{\bf Proactive defense via relay selection.} 
Sun \emph{et al.}~\cite{sun2017counter} proposed a new relay selection algorithm
to protect the connection between a Tor client and the entry relay. 
This algorithm defends against equally-specific prefix attacks on entry relays, 
where the effect is localized and only clients in certain locations will get affected
The localized effect opens up the possibility for clients to stay unaffected
by choosing the relay wisely and proactively before any attack happens. 
The algorithm maximizes the probability of clients being unaffected by attacks based on the topological locations of the clients and the relays. It successfully improves the probability by 36\% on average (up to 166\% for certain Tor client locations).

{\bf Reactive defense via monitoring.} 
To complement the proactive defense, Sun \emph{et. al.} proposed
 a monitoring system on routing activities for Tor relays. 
The system uses new detection techniques such as time-based and frequency-based heuristics, specifically tuned for Tor. 
The authors showed that most BGP updates involving a Tor relay are only announced
by a single AS (across all updates), effectively differentiating the announcements made by adversary ASes who never announced the prefix in the past. 
Tan \emph{et al.}~\cite{tan2016data} also proposed a data-plane detection approach that periodically runs traceroute to
detect longest-prefix attacks and update Tor relay descriptors upon anomaly detection, so that Tor clients can pick entry
relays correspondingly.

\section{Certificate Authorities}
\label{sec:pki}

The Public Key Infrastructure is the foundation for securing online communications. Digital certificates are issued by trusted certificate authorities (CAs) to domain owners, verifying the ownership of a domain. Internet users trust a domain with encrypted communications, such as bank websites, only if a valid certificate signed by a CA is presented. This mechanism effectively prevents Man-In-The-Middle (MITM) attacks that can have disastrous consequences, such as stealing users' financial information. 

However, the certificate issuance process is itself vulnerable to routing attacks, allowing network-level adversaries to obtain trusted digital certificates for any victim domain~\cite{birge2018bamboozling}. 
These attacks have significant consequences for the integrity and privacy of online communications, as adversaries can use fraudulently obtained digital certificates to bypass the protection offered by encryption and launch man-in-the-middle attacks against critical communications.

\subsection{How Certificate Authorities Work}
\begin{figure}
\centering
\includegraphics[width=.9\linewidth]{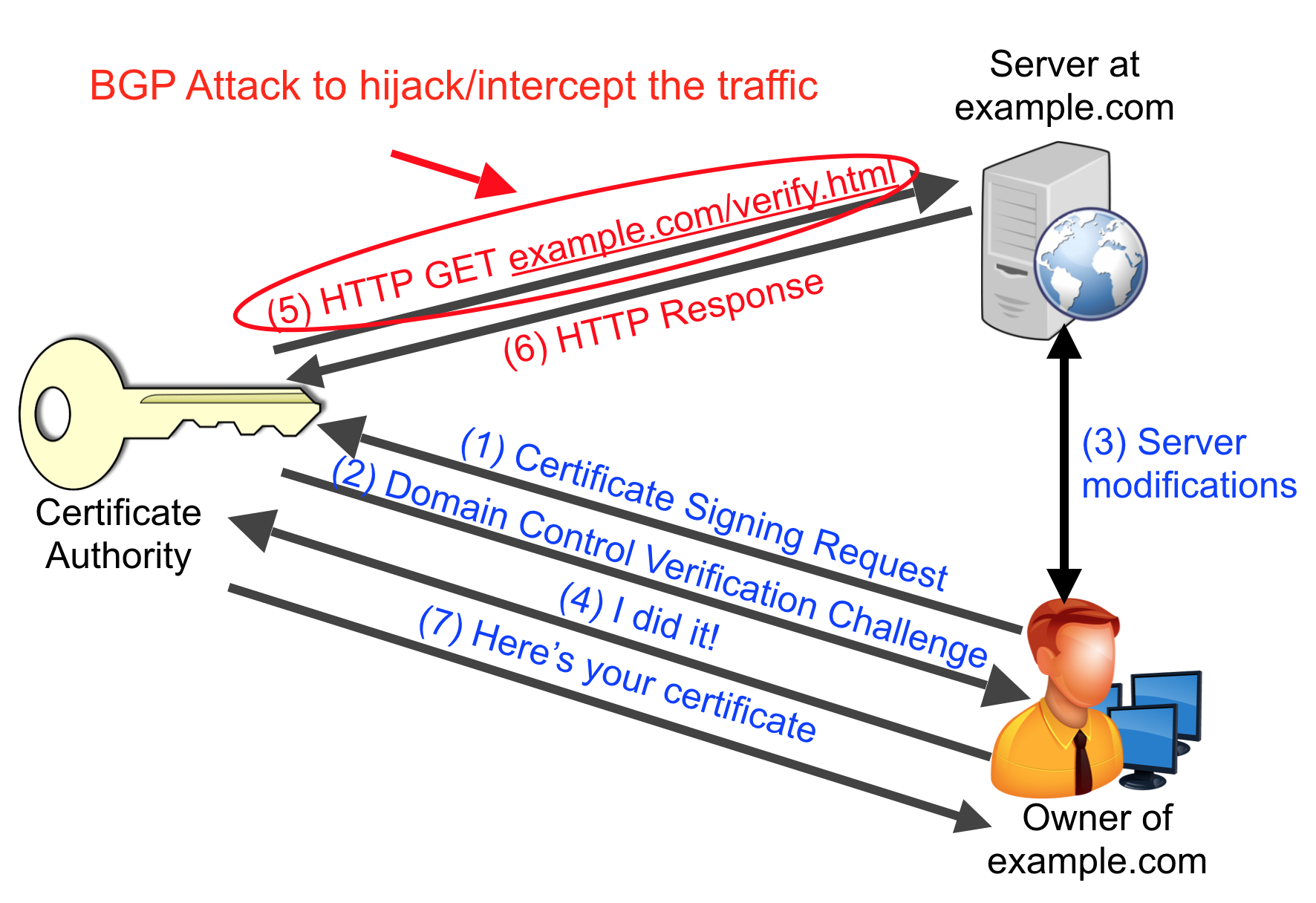}
\caption{BGP attack on domain control verification.}
\label{fig:hijack_dv}
\end{figure}

Domain control verification is a crucial process for domain owners to obtain digital certificates from CAs. Domain owners approach a CA %
to request a digital certificate, and the CA responds with a challenge that requires the owners to demonstrate control of an important network resource (e.g., a website or email address) associated with the domain. %
Figure~\ref{fig:hijack_dv} illustrates HTTP verification where 
the CA requires the domain owner to upload a document to a well-known directory on its web server and verify the upload over HTTP. 
Upon completion of the challenge, the CA issues the digital certificate to the domain owner. 

\subsection{Routing Attacks on Digital Certificates}

The domain control verification process creates a vulnerability to adversaries who can fake control of the network resources. 
Network-level adversaries can use routing attacks to hijack or intercept the traffic to the victim's domain such that the CA's request is routed to the adversaries instead~\cite{birge2018bamboozling} (step (5) in Figure \ref{fig:hijack_dv}). 
Adversaries can then answer the CA's HTTP request in step (6) and subsequently obtain a signed digital certificate from the CA for the victim domain. 
The attacks were successfully demonstrated in the real world, ethically~\cite{birge2018bamboozling}. 
The attacked domains were run on IP prefixes controlled by the researchers and had no real users or services.
The adversary successfully obtained certificates for the victim domain from five top CAs 
in as little as 35 seconds (Table \ref{tbl:CATable}). 

This work highlights the significant damage of routing attacks that can compromise the foundation of secure online communications, and shows the urgent need for practical defenses. 
Furthermore, the attacks also apply to other systems that require demonstration of control on certain resources via verification requests, such as email verifications. 
The communication with the mail server can be hijacked or intercepted, and there is still a non-negligible amount of emails that are unencrypted (e.g., less than 20\% of the emails from ``icicibank.com", a bank website, are encrypted~\cite{google-transparency}). 

\begin{table}[!t]
\scriptsize
\begin{center}
 \begin{tabular}{|p{1.4cm}|p{1.0cm}|p{1.1cm}|p{1.0cm}|p{1.05cm}|p{1.15cm}|}
 \hline
 &Let's Encrypt&GoDaddy&Comodo&Symantec&GlobalSign\\
 \hline
 \textbf{Time to issue certificate}&35s&$<$10min&51s&6min&4min\\
 \hline
 \textbf{Human Interaction}&No&No&No&No&No\\
 \hline
\textbf{Multiple Vantage Points}&No\footnote{No vantage points were deployed at time of attack. Let's Encrypt has since deployed multiple vantage point verification~\cite{letsencrypt2020multiva}.}&No&No&No&No\\
 \hline
\textbf{Validation Method Attacked}&HTTP&HTTP&Email&Email&Email\\
\hline
 \end{tabular}
\caption{Five CAs were attacked and obtained certificates from. All were automated and none had any defenses against BGP attacks.}
\label{tbl:CATable}
\end{center}
\vspace{-.2in}
\end{table}

\subsection{Defenses to Protect Digital Certificates}

Many currently deployed defenses do not sufficiently protect digital certificates. Given the relatively  short time required to obtain a fraudulent certificate, adversaries can get a certificate before the attack is mitigated, even if it is detected by monitoring systems. In addition, adversaries can potentially obtain a malicious certificate using only localized routing attacks that do not affect a large portion of the Internet. If a domain does not have a CAA DNS record (which is currently true of the vast majority of domains~\cite{CAA_study}), any CA is authorized to sign a certificate for that domain. Thus, adversaries only need to affect the route between one (of several hundred) CAs and the target domain to obtain a fraudulent certificate.

Birge-Lee \emph{et al.}~\cite{birge2018bamboozling} recently proposed two practical application-layer defenses. (1) Multiple Vantage Point Verification: building on the key insight that routing attacks may be localized, CAs can significantly decrease their vulnerability to attacks by performing domain verification from multiple vantage points and suspend certificate issuance in the case of inconsistent validation results. By adding only one additional vantage point, the probability of catching a localized routing attack on a domain increases from 61\% to 84\%. By having two additional vantage points, the probability of catching the attack reaches over 90\% for 74\% of the 1.8 million domains in the study. 
(2) BGP monitoring with route age heuristics: building on the key insight that anomalous and suspicious routing announcements are usually short-lived, CAs can require the routes to the domains to be active for a minimum time threshold before signing a certificate. 
This defense would force attacks to be active for over a day before the routes can be used to obtain a bogus certificate. 
Both defenses only require minimal deployment effort by the CAs with no change needed from domain owners or the routing infrastructure.

Multiple vantage point verification has gained significant traction. Let's Encrypt, the world's largest CA, has deployed multiple vantage point verification~\cite{letsencrypt2020multiva}. Furthermore, the prominent CDN CloudFlare has developed an API for CAs to perform multiple vantage point verification using its network~\cite{cloudflaremultipath}. %

\section{The Bitcoin Network}
\label{sec:bitcoin}

Bitcoin is the most widely-used cryptocurrency to date with over 42 million users~\cite{many_clients}.  
However, network-level adversaries can launch routing attacks to partition the bitcoin network, effectively preventing the system from reaching consensus~\cite{apostolaki2017hijacking}. 
Besides Bitcoin, this attack is generally applicable to many peer-to-peer networks and is particularly dangerous against blockchain systems. %

\begin{figure*}
  \centering
  \begin{subfigure}[b]{0.45\textwidth} 
   \centering 
   \includegraphics[width=\textwidth]{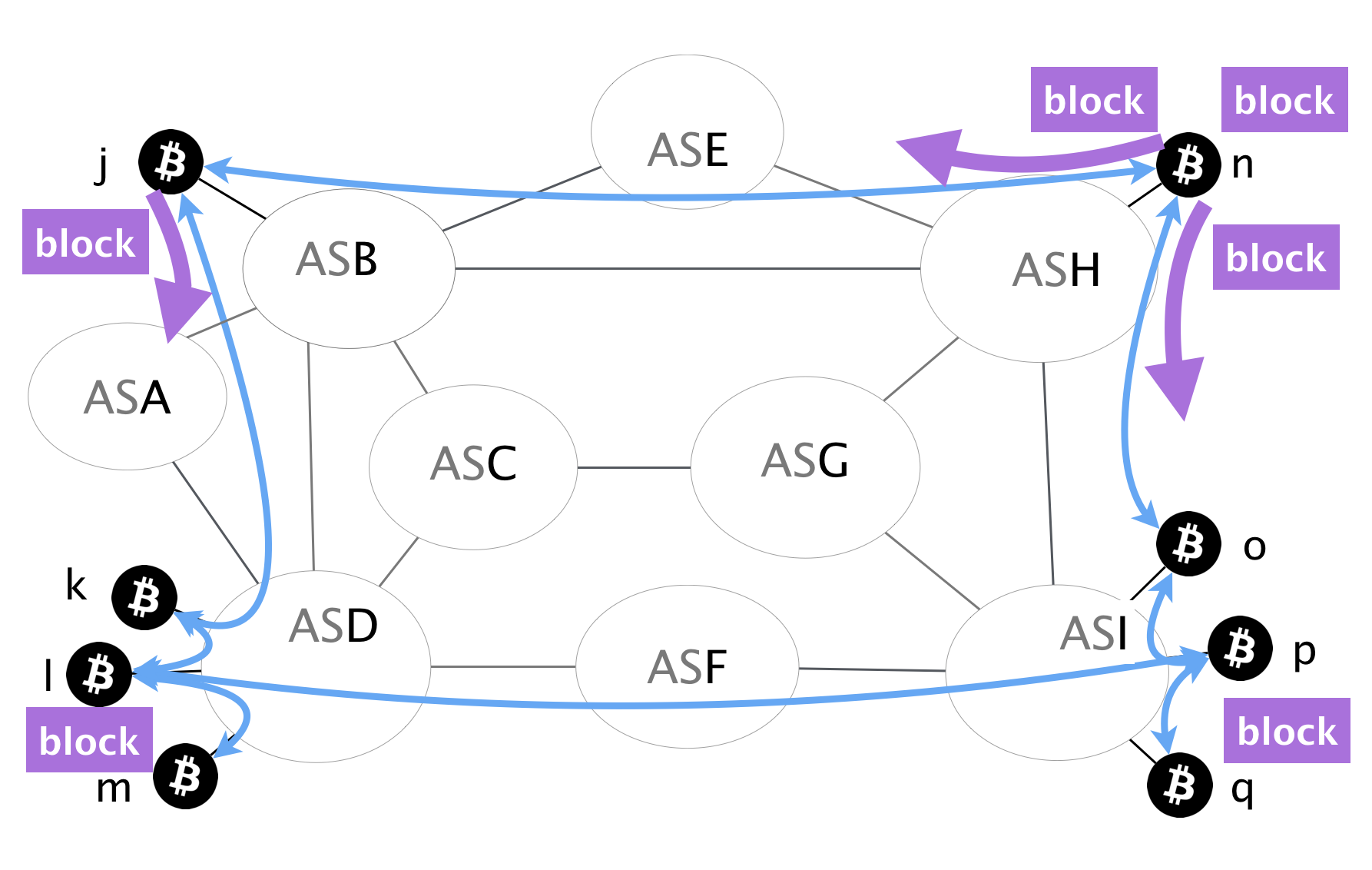}
   \caption[]%
   {{\small  New blocks mined by bitcoin nodes in different ASes are propagated to the whole network.}}
   \label{fig:part2}
  \end{subfigure}
    \hfill
   \begin{subfigure}[b]{0.45\textwidth} 
   \centering 
   \includegraphics[width=\textwidth]{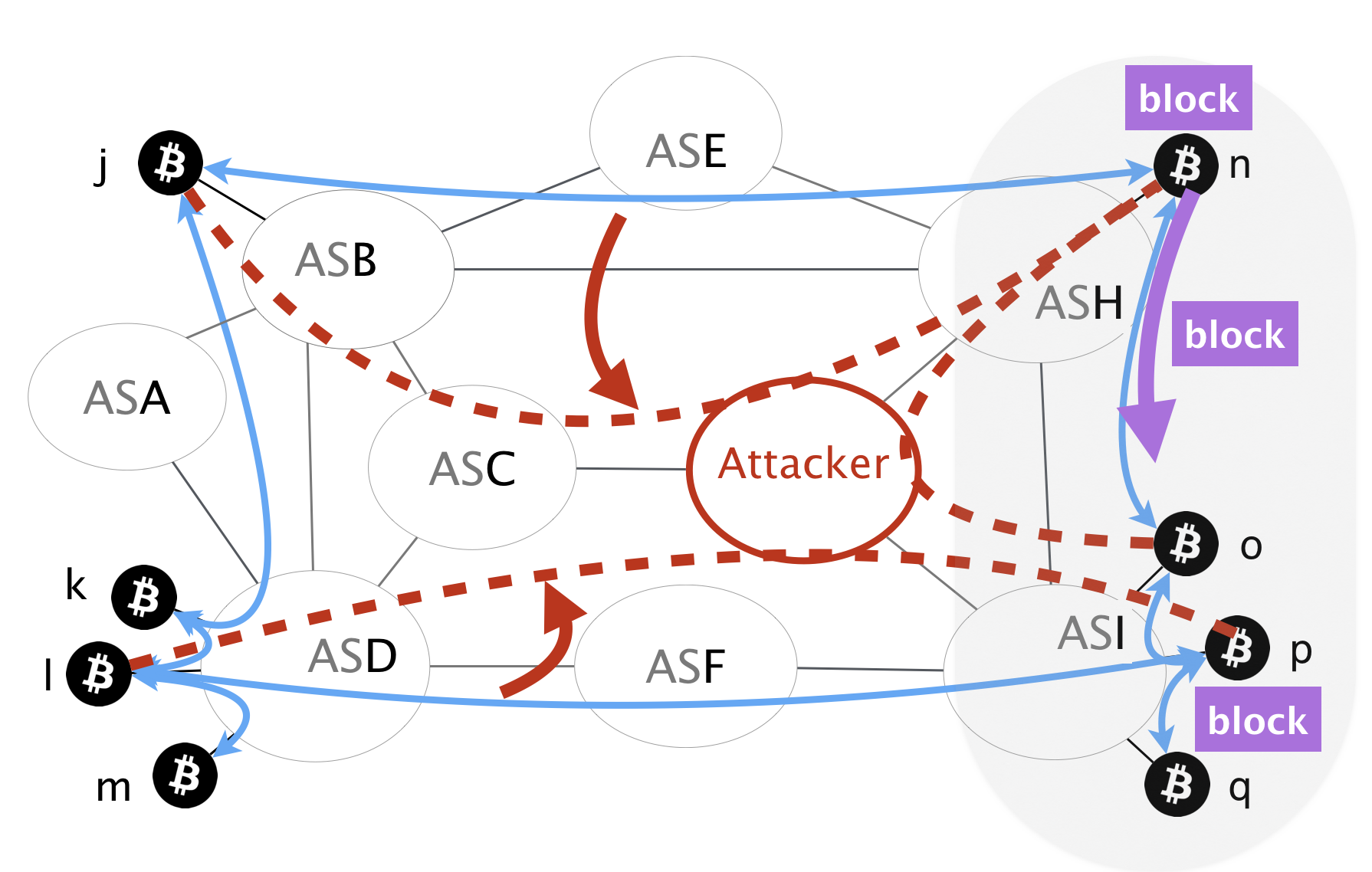}
   \caption[]%
   {{\small The attacker hijacks all prefixes pertaining to bitcoin nodes in the gray zone. Consequently, blocks mined by nodes in the gray zone won't be propagated further, which effectively isolates the gray zone.}}
   \label{fig:part3}
  \end{subfigure}
\caption[]{}
\end{figure*}

\subsection{How Bitcoin Works}
\label{sec:btc}
Bitcoin is a peer-to-peer network in which nodes use consensus mechanisms to jointly agree on a (distributed) log of all the transactions that ever happened. This log is called the \emph{blockchain} because it is composed of an ordered list (chain) of grouped transactions (blocks).

Special nodes, known as wallets, are responsible for originating transactions and propagating them in the network using a gossip protocol. %
A different set of nodes, known as miners,
are responsible for verifying the most recent transactions, grouping them in a block, and appending this block to the blockchain. To do so, the miners need to solve a periodic puzzle whose complexity is automatically adapted to the computational power of the miners in the network. %

Every time a miner creates a block, it broadcasts it to all the nodes in the network and receives freshly-mined bitcoins. Besides the most recent transactions, the block contains a proof-of-work (a solution to the puzzle) that each node can independently verify before propagating the block further.
In Figure~\ref{fig:part2}, node $n$ ``mines'' a block which is then broadcasted hop-by-hop in the network. %

As miners work concurrently, several of them may find a block at nearly the same time. %
These blocks effectively create ``forks'' in the blockchain, i.e., different
versions of the blockchain. The conflicts are eventually resolved as
subsequent blocks are appended to each chain and one of them becomes longer. In
this case, the network automatically discards the shorter chains, effectively
discarding the corresponding blocks together with the miner's revenues.

\subsection{Routing Attacks on Consensus}
\label{sec:partition}

Network-level adversaries can perform routing attacks on bitcoin to partition the set of nodes into two (or more) disjoint components~\cite{apostolaki2017hijacking}. Consequently, the attacks disrupt the ability of the entire network to reach consensus.
The adversary must divert and cut \emph{all} the connections connecting the various components together.
To do so, the adversary can perform an interception attack by hijacking the IP prefixes of each component and selectively dropping the connections crossing the components, while leaving the internal connections (within a component) untouched. 

In Figure~\ref{fig:part3}, the adversary hijacks all prefixes pertaining to bitcoin nodes in the gray zone. 
Having gained control over the traffic towards these nodes (red lines), the adversary drops the connections between the clients that are within the gray zone and outside it, effectively creating a partition. %

The impact of partition attacks is worrying. First, a partition attack can act as a denial-of-service attack: clients can neither properly propagate the corresponding transactions, nor verify the ownership of funds. Second, a partition attack can lead to high revenue loss for the miners: once the network reconnects, the shortest chain(s) will be discarded, permanently depriving miners of their rewards.

\subsection{Defenses to Protect the Bitcoin Consensus}
\label{sec:relay}
\begin{figure*}
  \centering
  \begin{subfigure}[b]{0.45\textwidth} 
   \centering 
   \includegraphics[width=\textwidth]{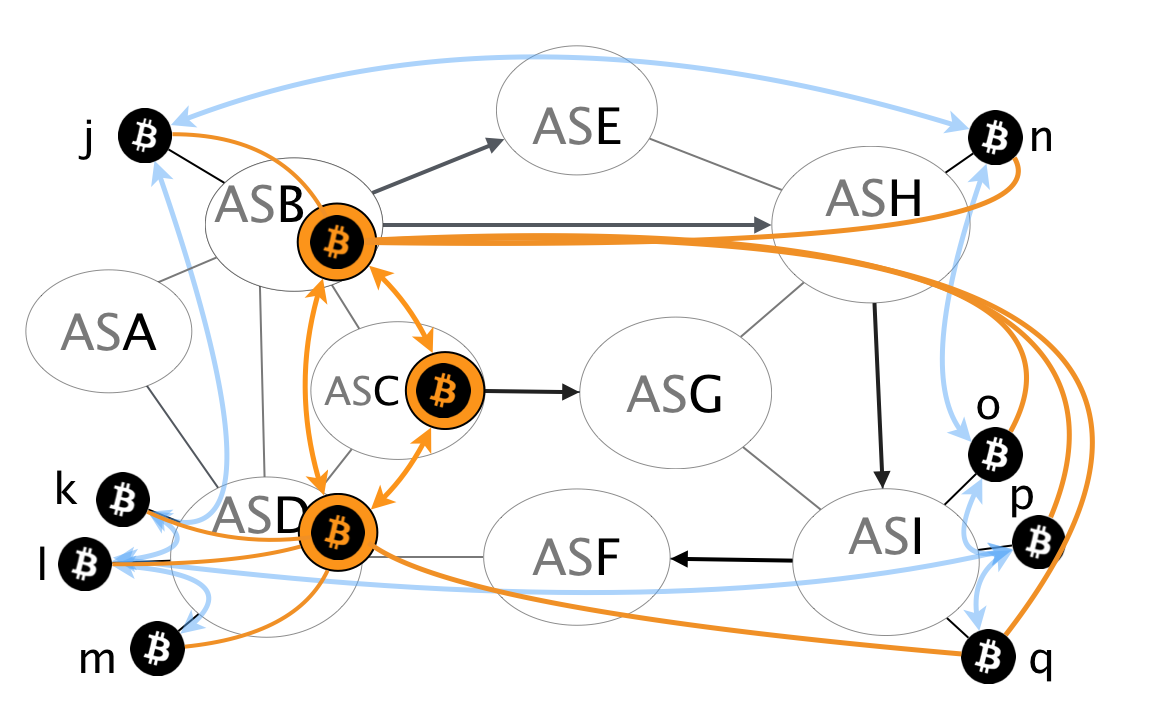}
   \caption[]%
   {{\small Relay nodes are hosted in ASes that have no customer ASes and compose connected graph of direct peering links. Bitcoin clients connect to at least one relay node. } } 
   \label{fig:part5}
  \end{subfigure}
    \hfill
   \begin{subfigure}[b]{0.45\textwidth} 
   \centering 
   \includegraphics[width=\textwidth]{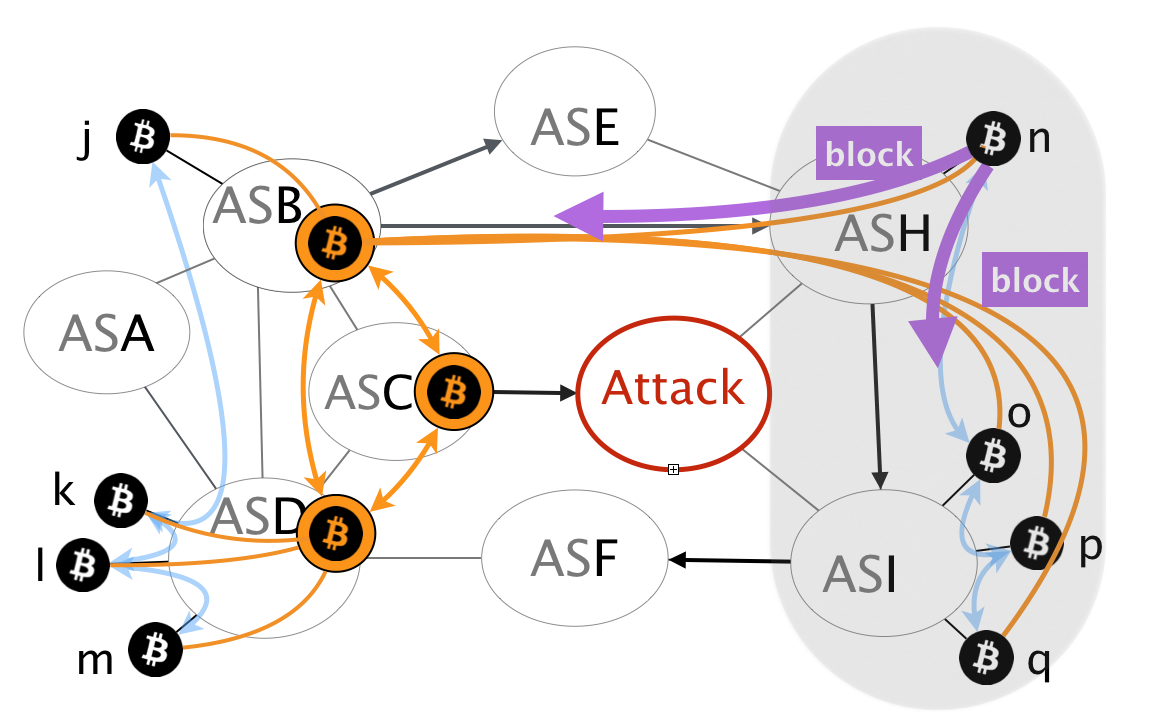}
   \caption[]%
   {{\small While routing attacks isolate the gray zone from the rest of the bitcoin network, blocks mined in the gray zone are propagated via relay nodes in the overlay SABRE network. }} 
   \label{fig:part6}
  \end{subfigure}
   \label{fig:partall}
\caption[]{}
\end{figure*}

Apostolaki \emph{et al.}~\cite{ApostolakiMMV19} recently proposed SABRE to protect bitcoin from partition attacks.
SABRE is an overlay network, composed of a small set of special bitcoin clients (relays) that receive, verify, and propagate blocks. Regular bitcoin clients can connect to one or more relays in addition to their regular connections. %
During a partition attack, SABRE relays stay connected to each other and to many bitcoin clients, allowing block propagation among the otherwise disconnected components. 
In Figure~\ref{fig:part6}, while clients in the gray zone are isolated from the rest of the network, a block mined by node $n$ is propagated via the relay nodes (colored in orange) to the rest of the network. 

SABRE achieves this by strategically choosing the ASes in which to host relay
nodes. %
The key insight is that some ASes,
such as those without customers, are naturally protected against routing
attacks. By hosting relays in these ASes, SABRE can therefore maintain its
connectivity and its ability to propagate blocks on behalf of bitcoin clients,
even in the presence of routing attacks. Note that a bitcoin client only
requires one unhindered connection to a SABRE relay to be protected.

In the SABRE network shown in Figure~\ref{fig:part5}, three ASes ($ASB$, $ASC$, $ASD$) are selected to host the relay nodes, which directly peer with each other and have no customer ASes. 
During routing attacks, the relay nodes stay connected to each other. 
For instance, if $ASG$ (provider of $ASC$) announces the prefix of $ASB$, $ASC$ would still prefer the route to $ASB$ since it's via a peer. 
Additionally, all bitcoin clients keep at least one connection to the relay network during the attack. 
Even nodes such as node $q$ which loses one of the connections to the relay network due to the attack, stays connected via another relay node.

 \section{Cross-layer Solutions}
\label{sec:discussion}

We demonstrated the emerging threats of routing attacks to critical applications. Next, we outline lessons learned from the three applications, and discuss the importance of developing solutions at both the application and network layers. 

\subsection{For Application Developers}

The most important takeaway is the significant impact of routing (in)security on Internet applications. 
When securing the application layer in isolation becomes hard to achieve, we should think about cross-layer solutions that take into account routing properties at the network layer. 

\begin{figure}
\centering
\includegraphics[width=.9\linewidth]{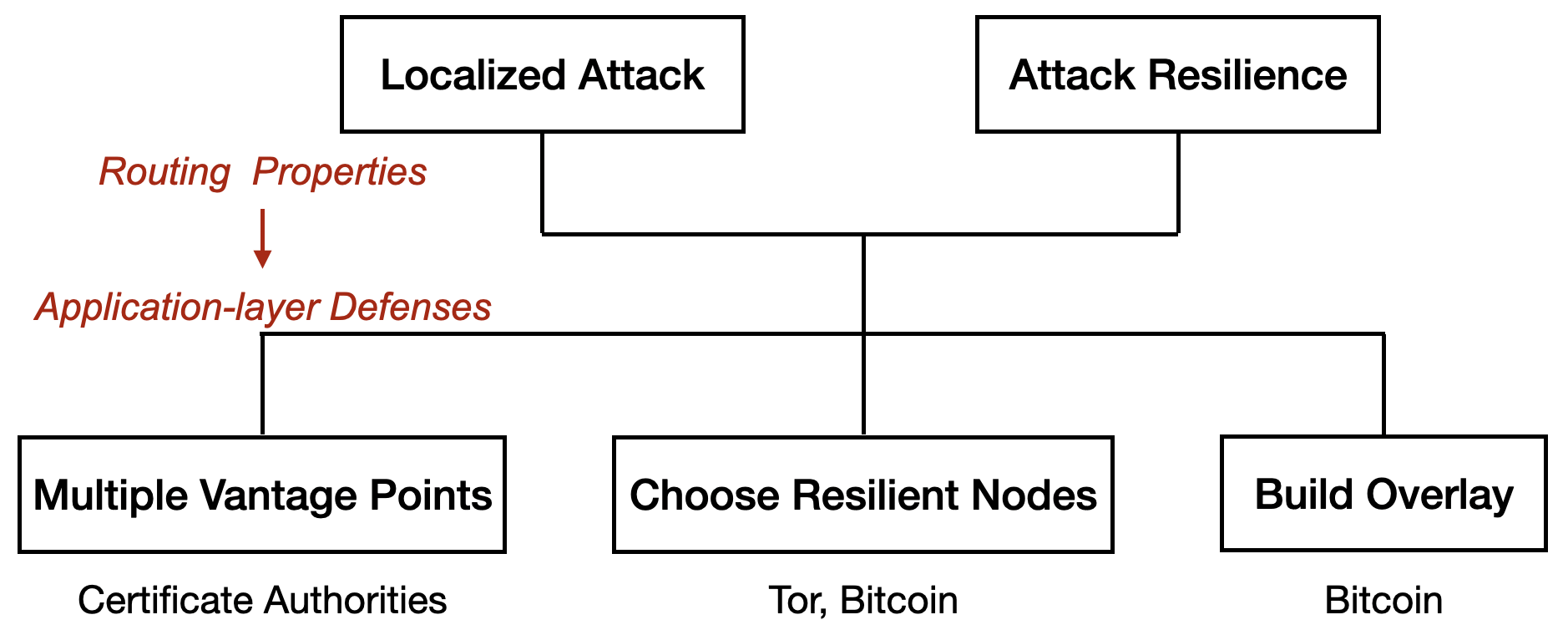}
\caption{Two routing properties serve as the key insights in developing application-layer defenses.}
\label{fig:discussion}
\end{figure}

We outline two routing properties that are the key insights in building application-layer defenses:
(i) \emph{localized attack}: attack announcements may not be propagated and visible to the whole Internet, and stealthy adversaries can carefully craft announcements to control propagation and only target certain regions; (ii) \emph{attack resilience}: some ASes that receive the attack announcement may not be affected, i.e., not favoring the malicious path and hence being ``resilient" to the attack. %
This depends on the routing preferences, e.g., if an AS receives the attack announcement from a provider while the legitimate path is through a peer, the AS will still prefer the legitimate path. %

These two simple routing properties lead to three generalizable application-layer defenses shown in Figure~\ref{fig:discussion}. 

{\bf Deploy multiple vantage points.} Initiating connections from multiple vantage points increases the likelihood of detecting and circumventing a localized attack. Certificate Authorities can perform domain control verification from multiple vantage points to ensure that routes to the destination are consistent. This approach generalizes to a broad set of verification processes, where verifications from multiple sources would help lower the success of an attack and significantly increase the cost to an adversary. BGP monitoring systems also benefit from having more comprehensive data through multiple vantage points to detect stealthy attack. 

{\bf Choose resilient nodes.} Applications can strategically choose nodes/servers that are the most resilient to attacks. Tor clients may choose an entry relay that maximizes the probability of being resilient given the AS locations of the client and the relay. Bitcoin may choose relay nodes in certain ASes (e.g., peer AS without customers) to avoid being affected by attacks. The specific implementation can vary based on the need of the applications, and may even bring in RPKI as a criteria in choosing resilient nodes. 

{\bf Build an overlay network.} This approach can help mitigate some effects of routing attacks, e.g., partitioning Bitcoin nodes, by providing alternative routes. It can be more effective when combined with ``choosing resilient nodes", where the nodes in the overlay are carefully chosen to maximize the resilience to attacks.
Bitcoin is an example application that benefits from an overlay to mitigate partitioning attacks, but the approach is generally applicable to many peer-to-peer networks. 

\subsection{For Network Operators} 

While application-layer defenses can provide immediate protections, we should also push for large-scale deployment of general defenses against sophisticated routing attacks.  We recommend that ASes (i) adopt best practices outlined in the MANRS~\cite{manrs} project, (ii) accelerate the adoption of RPKI by publishing ROAs and performing Route Origin Validation (ROV), and (iii) build consensus on a pathway to solving routing security issues (including full path security) once and for all. Furthermore, we outline two ways that synergize network operators with application developers. 

{\bf Applications as starting points.} Securing all 800K prefixes and 67K ASes~\cite{cidr-report} seems like an impossible task.  However, only a small portion of the prefixes play a heavy role in each application.  For instance, only around $1100$ ASes have Tor relays hosted on their prefixes, and one AS alone carries 23\% of all Tor traffic~\cite{sun2017counter}.  Furthermore, in digital certificate issuance, a handful of certificate authorities issue the vast majority of certificates, and the domains are largely hosted on a few cloud and CDN providers (e.g., five ASes including SquareSpace and Amazon host nearly half of the domains~\cite{birge2018bamboozling}). Finally, only 5 ASes host one third of all Bitcoin clients~\cite{Bitnodes}, while 50\% of all mining power is hosted in less that 100 prefixes~\cite{apostolaki2017hijacking}. If a few thousand ASes can take major steps to deploy routing security, the applications will receive tremendous benefits.

{\bf Applications as incentives.} Popular applications---and their users---can help incentivize the deployment of routing security solutions by the actions they take, while ensuring the applications' security/privacy goals.  For instance, Tor could favor certain relays that are hosted on authenticated prefixes, and domain owners could favor cloud hosting services that provide origin validations and favor certificate authorities hosted on authenticated prefixes. Similarly, miners could prefer hosting their infrastructure in ASes that provide origin validation, while regular client could prefer to connect to peers hosted on authenticated prefixes. These steps may help motivate network operators to validate their prefixes to offer better service to their customers, and eventually lead to a more secure routing infrastructure.

\section{Conclusion}
\label{sec:conclusion}

Often times, we focus on individual system layers in isolation. 
In neglecting routing (in)security, application developers underestimate the risks for their users. In focusing on availability threats, network operators underestimate the risks to Internet applications. By demonstrating the dire consequences of routing attacks on Internet applications, we stress the importance of cross-layer awareness and the need to deploy both application-layer and network-layer solutions.

\bibliographystyle{abbrv} %
\bibliography{main}  %

\begin{thebibliography}{10}

\bibitem{Bitnodes}
{Bitnodes}.
\newblock \url{https://bitnodes.io/dashboard/}.

\bibitem{google-transparency}
{Google Transparency Report}.
\newblock \url{https://transparencyreport.google.com/safer-email/}.

\bibitem{nsashaping}
{Network Shaping 101}.
\newblock
  \url{https://www.documentcloud.org/documents/2919677-Network-Shaping-101.html}.

\bibitem{nist-rpki}
{RPKI Deployment Monitor}.
\newblock \url{https://rpki-monitor.antd.nist.gov/}.

\bibitem{tormetrics}
Tor metrics.
\newblock \url{https://metrics.torproject.org/}.

\bibitem{youtube2008}
{Pakistan hijacks YouTube}.
\newblock \url{https://dyn.com/blog/pakistan-hijacks-youtube-1/}, 2008.

\bibitem{google2017}
{Google leaked prefixes -- and knocked Japan off the Internet}.
\newblock
  \url{https://www.internetsociety.org/blog/2017/08/google-leaked-prefixes-knocked-japan-off-internet/},
  2017.

\bibitem{etherwaller}
{AWS DNS network hijack turns MyEtherWallet into ThievesEtherWallet}.
\newblock
  \url{https://www.theregister.co.uk/2018/04/24/myetherwallet_dns_hijack/},
  2018.

\bibitem{bitcanal2018}
{Shutting Down the BGP Hijack Factory}.
\newblock
  \url{https://blogs.oracle.com/internetintelligence/shutting-down-the-bgp-hijack-factory},
  2018.

\bibitem{chinatelecom2019}
{BGP event sends European mobile traffic through China Telecom for 2 hours}.
\newblock
  \url{https://arstechnica.com/information-technology/2019/06/bgp-mishap-sends-european-mobile-traffic-through-china-telecom-for-2-hours/},
  2019.

\bibitem{cloudflaremultipath}
{Securing Certificate Issuance using Multipath Domain Control Validation}.
\newblock \url{https://blog.cloudflare.com/secure-certificate-issuance/}, 2019.

\bibitem{cidr-report}
{CIDR Report}.
\newblock \url{http://www.cidr-report.org/as2.0/}, 2020.

\bibitem{manrs}
{MANRS Project}.
\newblock \url{https://www.manrs.org/}, 2020.

\bibitem{letsencrypt2020multiva}
{Multi-Perspective Validation Improves Domain Validation Security}.
\newblock
  \url{https://letsencrypt.org/2020/02/19/multi-perspective-validation.html},
  2020.

\bibitem{many_clients}
{Number of Blockchain wallet users worldwide}.
\newblock
  \url{https://www.statista.com/statistics/647374/worldwide-blockchain-wallet-users/},
  2020.

\bibitem{ApostolakiMMV19}
M.~Apostolaki, G.~Marti, J.~M{\"{u}}ller, and L.~Vanbever.
\newblock {SABRE: Protecting Bitcoin against Routing Attacks}.
\newblock In {\em Network and Distributed System Security Symposium (NDSS)},
  2019.

\bibitem{apostolaki2017hijacking}
M.~Apostolaki, A.~Zohar, and L.~Vanbever.
\newblock {Hijacking Bitcoin: Routing Attacks on Cryptocurrencies}.
\newblock In {\em IEEE Symposium on Security and Privacy}, 2017.

\bibitem{birge2018bamboozling}
H.~Birge-Lee, Y.~Sun, A.~Edmundson, J.~Rexford, and P.~Mittal.
\newblock {Bamboozling Certificate Authorities with BGP}.
\newblock In {\em USENIX Security Symposium}, 2018.

\bibitem{birge2019sico}
H.~Birge-Lee, L.~Wang, J.~Rexford, and P.~Mittal.
\newblock {SICO: Surgical Interception Attacks by Manipulating BGP
  Communities}.
\newblock In {\em ACM Conference on Computer and Communications Security
  (CCS)}, 2019.

\bibitem{boldyreva2012provable}
A.~Boldyreva and R.~Lychev.
\newblock {Provable security of S-BGP and other path vector protocols: Model,
  analysis and extensions}.
\newblock In {\em ACM Conference on Computer and Communications Security
  (CCS)}, 2012.

\bibitem{rpki}
R.~Bush and R.~Austein.
\newblock {The Resource Public Key Infrastructure (RPKI) to Router Protocol}.
\newblock RFC 6810, RFC Editor, January 2013.

\bibitem{dingledine2004tor}
R.~Dingledine, N.~Mathewson, and P.~Syverson.
\newblock Tor: the second-generation onion router.
\newblock In {\em USENIX Security Symposium}, 2004.

\bibitem{gill2011let}
P.~Gill, M.~Schapira, and S.~Goldberg.
\newblock {Let the market drive deployment: A strategy for transitioning to BGP
  security}.
\newblock In {\em ACM SIGCOMM}, 2011.

\bibitem{goldberg2017nsa}
S.~Goldberg.
\newblock {Surveillance without borders: The ``traffic shaping" loophole and
  why it matters}.
\newblock {\em The Century Foundation}, 2017.

\bibitem{hu2007accurate}
X.~Hu and Z.~M. Mao.
\newblock Accurate real-time identification of {IP} prefix hijacking.
\newblock In {\em IEEE Symposium on Security and Privacy}, 2007.

\bibitem{kent2000secure}
S.~Kent, C.~Lynn, and K.~Seo.
\newblock {Secure border gateway protocol (S-BGP)}.
\newblock {\em IEEE Journal on Selected areas in Communications},
  18(4):582--592, 2000.

\bibitem{lad2006phas}
M.~Lad, D.~Massey, D.~Pei, Y.~Wu, B.~Zhang, and L.~Zhang.
\newblock {PHAS}: A prefix hijack alert system.
\newblock In {\em USENIX Security Symposium}, 2006.

\bibitem{bgpsec}
M.~Lepinski and K.~Sriram.
\newblock {BGPsec Protocol Specification}.
\newblock RFC 8205, RFC Editor, September 2017.

\bibitem{lychev2013bgp}
R.~Lychev, S.~Goldberg, and M.~Schapira.
\newblock {BGP security in partial deployment: Is the juice worth the squeeze?}
\newblock In {\em ACM SIGCOMM}, 2013.

\bibitem{qiu2007detecting}
J.~Qiu, L.~Gao, S.~Ranjan, and A.~Nucci.
\newblock Detecting bogus {BGP} route information: Going beyond prefix
  hijacking.
\newblock In {\em SecureComm}, 2007.

\bibitem{reuter2018towards}
A.~Reuter, R.~Bush, I.~Cunha, E.~Katz-Bassett, T.~C. Schmidt, and
  M.~W{\"a}hlisch.
\newblock {Towards a rigorous methodology for measuring adoption of RPKI route
  validation and filtering}.
\newblock {\em ACM SIGCOMM Computer Communication Review}, 48(1):19--27, 2018.

\bibitem{CAA_study}
Q.~Scheitle, T.~Chung, J.~Hiller, O.~Gasser, J.~Naab, R.~van Rijswijk-Deij,
  O.~Hohlfeld, R.~Holz, D.~Choffnes, A.~Mislove, and G.~Carle.
\newblock A first look at certification authority authorization {(CAA)}.
\newblock {\em SIGCOMM Comput. Commun. Rev.}, 48(2):10–23, May 2018.

\bibitem{peering19}
B.~Schlinker, T.~Arnold, I.~Cunha, and E.~Katz-Bassett.
\newblock {PEERING}: Virtualizing {BGP} at the edge for research.
\newblock In {\em ACM SIGCOMM CoNEXT Conference}, Dec. 2019.

\bibitem{shi2012detecting}
X.~Shi, Y.~Xiang, Z.~Wang, X.~Yin, and J.~Wu.
\newblock Detecting prefix hijackings in the {Internet} with {Argus}.
\newblock In {\em Internet Measurement Conference (IMC)}, 2012.

\bibitem{peerlocking}
J.~Snijders.
\newblock {Practical everyday BGP filtering with AS\_PATH filters:
  PeerLocking}.
\newblock {\em NANOG-67, Chicago, June}, 2016.

\bibitem{sun2017counter}
Y.~Sun, A.~Edmundson, N.~Feamster, M.~Chiang, and P.~Mittal.
\newblock {Counter-RAPTOR: Safeguarding Tor against Active Routing Attacks}.
\newblock In {\em IEEE Symposium on Security and Privacy}, 2017.

\bibitem{sun2015raptor}
Y.~Sun, A.~Edmundson, L.~Vanbever, O.~Li, J.~Rexford, M.~Chiang, and P.~Mittal.
\newblock {RAPTOR: Routing Attacks on Privacy in Tor}.
\newblock In {\em USENIX Security Symposium}, 2015.

\bibitem{tan2016data}
H.~Tan, M.~Sherr, and W.~Zhou.
\newblock Data-plane defenses against routing attacks on {Tor}.
\newblock In {\em Privacy Enhancing Technologies Symposium (PETS)}, 2016.

\bibitem{zhang2008ispy}
Z.~Zhang, Y.~Zhang, Y.~C. Hu, Z.~M. Mao, and R.~Bush.
\newblock {iSpy: Detecting IP prefix hijacking on my own}.
\newblock In {\em ACM SIGCOMM}, 2008.

\bibitem{zheng2007light}
C.~Zheng, L.~Ji, D.~Pei, J.~Wang, and P.~Francis.
\newblock A light-weight distributed scheme for detecting {IP} prefix hijacks
  in real-time.
\newblock In {\em ACM SIGCOMM}, 2007.

\end{thebibliography}
\balancecolumns
\end{document}